# Mechanical tunning of adhesion through micro-patterning of elastic surfaces

Christophe Poulard,*[a] Frédéric Restagno,[a] Raphaël Weil,[a] and Liliane Léger[a]

We present an investigation of the role of micropatterning on adhesion properties at soft deformable polydimethylsiloxane (PDMS)/ acrylic adhesive interfaces. Contrary to what has been observed for low aspect ratio rigid patterns, where the adhesion enhancement was found to only result from the increase of interfacial area due to patterning, we show that for soft elastic arrays of cylindrical pillars, the elastic deformation of the patterns can lead to noticeble extra adhesion increase. The effect of the geometrical characteristics of the patterning for hexagonal arrays of PDMS micropillars on the adhesion energy are presented. We show that varying the size of the pattern allows one to tune the adhesion energy, and that this adhesion enhancement saturates when the pillars become too close to each other, due a coupling of the elastic deformation fields inside the underlying substrate. A mechanical model has been developed and found in good quantitative agreement with experimental data, with as unique fitting parameter the rupture criteria for the adhesive on the top of the pillars. Such a rupture a rupture criteria can thus be extracted from systematic experiments on controlled patterned surfaces. This criteria remain sensitive to the chemistry of the surfaces.

## Introduction

The precise tuning of adhesion properties is of great interest for a number of applications, especially when one deals with weak adhesion[1-2]. Cross-linked polydimethylsiloxane (PDMS) substrates are commonly used in microfluidic systems or in "bio-inspired" surface fabrication[3-6] because of their low surface energy and weak chemical reactivity which provide anti-adhesive properties. For a number of other applications, such as protective layers of stickers for example, the adhesion of common acrylic adhesives is too low on PDMS substrates, which then need be formulated with additives reinforcing adhesion. A commonly followed path relies on chemical modification of the antiadhesive coating, which usually leads to adhesion enhancement that strongly depends on the chemical nature of the adhesive. The design of PDMS surfaces with tailored adhesive properties remains a real technical challenge. An alternative and potentially more universal solution based on microstructuration has been proposed and has started to be investigated recently by several authors[1-14]. From a theoretical point of view, Arzt et al.[10] have shown, using the Johnson Kendall Roberts (JKR) theory of adhesive contacts, that splitting up one contact into many smaller subcontacts increases adhesion and Persson et al.[11-12] have shown that the effective elastic modulus of a fibrillar structure is much smaller than that of the corresponding bulk material. As a result, a fibrillar structure is expected to be highly compliant, which should help in forming intimate contact. This is, of course, of fundamental importance for adhesion on both smooth and rough substrates. Later, Jagota et al.[13] have shown that the work required to separate a fibrillar structure from a substrate is larger than that of the same smooth material because the elastic strain energy stored in the fibrils, when they deform, is lost during pull-off. Hui et al.[14], for their part, have shown that in the case of fibrillar structures the stress concentration at the crack tip is redistributed over a zone described by a characteristic length significantly larger than the cross-sectional dimensions of the fibrils. Within this zone, the fibrils are under equal load-sharing conditions. Consequently, the failure of the interface involves a simultaneous failure of all fibrils inside this zone, which is quite different from the usual crack propagation for which stress concentration favors a sequential failure of fibrils starting with the fibrils closest to the crack tip.

A number of experiments have been conducted to examine how fibrillar structures could enhance adhesion. For example, Ghatak et al.[15] have studied the adhesion properties between an incision-patterned PDMS elastomer layer and a flexible plate. The authors mentioned that multiple crack arrest and initiation on such substrates should result in extra dissipation of the elastic energy as observed for the fracture of soft elastomers. If two-dimensional textured surfaces are used and if the length scale of these patterns is small enough (typically smaller than the decay length of the stress applied to the elastomer), then a large enhancement of the interfacial fracture toughness is observed. Geim et al.[16], using substrates made of polyimide hairs pillars supported by double-stick viscoelastic tape, obtained high pull-off forces. Similar experiments have been performed and analyzed by Hui et al.[17] They established that without the double-stick viscoelastic tape the adhesion is not increased by the fibrillar structure, suggesting that the enhanced adhesion measured by Geim et al.[16] was due to dissipation in the viscoelastic layer. Finally, Crosby et al.[18] have demonstrated through JKR experiments that adhesion between glass and PDMS substrates patterned by low aspect ratio cylindrical pillars could be altered from 20 to 400% of the value of conventional adhesion descriptors for nonpatterned interfaces. Different local separation processes at the interface were observed, and general relationships between material properties, pattern length scales, and adhesion were established, depending on the characteristic sizes of the array (typical value are a few micrometers for the height and 20 to 500μm for the pillar radius and edge-to-edge spacing).

All of these findings are indeed along the lines of early experiments from Fuller and Tabor[19], who measured the rolling resistance of unvulcanized rubber on a rigid substrate. They observed that adhesion was enhanced on rough substrates, a result interpreted using two major assumptions:
- The viscoelastic properties of the rubber are essential to forming an intimate contact with a rough surface (as a result of stress relaxation).
- The substrate roughness leads to the formation of isolated contact regions during peeling.

The final stages of separation involve only isolated still-adhering zones whose associated elastic energy, built up during peeling, is lost when the contact is broken.

A first systematic comparison between soft elastic deformable patterns and rigid one, having exactly the same geometry and surface energy, and made to adhere to the same acrylic adhesive, has been reported recently[20]. By varying the geometry of the pillars, for an hexagonal array of micropillars with a spacing $i$, a height $h$ and a diameter $d$, two regimes of adhesion enhancement have been identified: for relatively low aspect ratio of the cylindrical pillars, ($h/d<1.5$), soft patterned substrates are more efficient than rigid ones in enhancing adhesion, pointing out the role of the deformation of the pattern; for pillars with higher aspect ratio, only rigid patterned surfaces do enhance adhesion. Then the only possible contribution to the energy dissipation comes from the enhanced viscoelastic losses in the adhesive layer. In the low aspect ration regime were adhesion enhancement is due to the deformation of the pillars, it was shown that the major mechanism to store elastic energy which is then lost after rupturing the contact can be either the deformation of the substrate locally pulled by the pillars ($h/d<0.8$) or the bending of the pillar ($h/d<0.8$).

It is thus now well admitted that fibrillar structures are involved in the formation of intimate contact with a rough substrates because they are more deformable than bulk smooth material and that different mechanisms of energy dissipation can lead to enhanced adhesion on fibrillar substrates either because of the stretching and deformation of the fibrils themselves or because of an overall modification of the stress and strain fields both inside the fibrillar zone and in the underlying material. These arguments, however, still remain qualitative, and the relative balance between these different contributions still has to be elucidated.

. In the present article, we report a systematic investigation of the exact role of the characteristic dimensions of the patterning on adhesion enhancement, focussing on the regime where the deformations in the elastic substrate are dominant, with a special emphasis on the effect of the coupling between the pattern structures when their relative distance is progressively decreased. To do so, PDMS substrates patterned with hexagonal arrays of cylindrical pillars with a fixed aspect ratio have been used. The diameter and the spacing of pillars have been systematically varied in order to gain deeper fundamental insight into the role of patterning on adhesion by identifying the relevant parameters of the pattern that control the modulation of adhesion.

## Experimental setup

The peel force $F$ was measured with a lab-developed peel apparatus schematically presented in figure 1. The top part includes a force sensor fixed on a 45° motorized endless screw which allows 90° peeling at a velocity $V$ in the range 0.5-5000 μm/s.

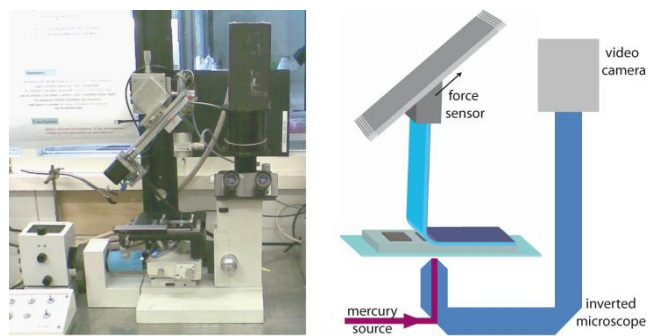

**Fig.1:** Picture and schematic representation of the peel experiment. Samples are put onto an inverted microscope and the peel motion is imposed by a 45° motorized endless screw coupled to motor. The peel force is measured by a force sensor attached to the peeled ribbon.

The bottom part is an inverted optical microscope allowing to visualize the peel front at micrometric scale. The sample is illuminated perpendicularly to its plane with a mercury source, through an optical fiber. Prior to each experiment, the adhesive tape (3M 600) was put into contact with the substrate, under a load corresponding to a pressure of 0.1 MPa, for 12 hours to ensure that the adhesive fully fills the space between the micropatterns (which can easily be checked optically, due to the refractive index matching when intimate contact between the substrate and the adhesive is attained).

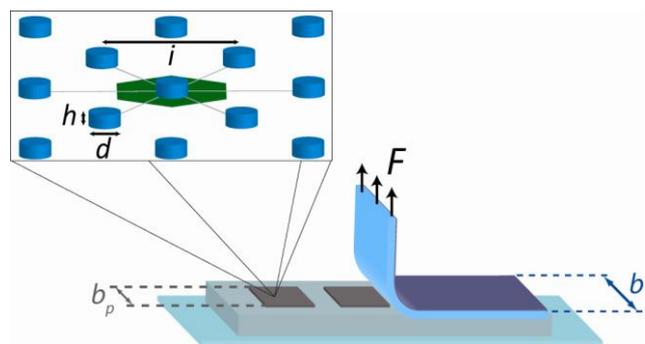

**Fig.2:** Schematic view of the geometry of a peel sample: the acrylic adhesive in contact with a patterned substrate is peeled at 90°. The width of the contact $b$ is 19 mm and the width of the patterned area $b_p$ is 8 mm.

Patterned PDMS substrates were produced by classical molding techniques using a silicon wafer with an etched resist layer as a mold. This mold was obtained with standard electronic lithography techniques: a thin layer of a positive resist (MicroChem PMMA 950k) is spincoated onto a silicon wafer and its thickness fixes the height *h* of the pillars. This resist layer was then locally exposed to an electron beam (30kV, 13nA) in a FESEM (Zeiss SUPRA 55VP). The desired pattern was first design with DesignCAD Express V16.2 and the FESEM was controlled with NPGS V9.0.190 to write the pattern. After irradiation, the exposed parts of the resists was develop in MIBK:IPA solution (3:1) during 60 sec under agitation. PDMS replicas were obtained by pouring in this mold a millimeter thick layer of PDMS mixed with a crosslinker (Sylgard 184, Dow Corning), curing at 50 °C for 24 h, and finally peeling off the crosslinked PDMS elastomer from the mold. The patterned PDMS elastomer films were finally fixed on a pretreated (UV-Ozone) glass plate. The elastic modulus of the films, E=1.0±0.1 MPa has been measured by a JKR test[21]. On a typical elastomer film, three succesive zones (8 mm by 8 mm) were patterned, and separated by non patterned zones (5 mm), as shown schematically in figure 2, so that the peel force on both smooth and patterned surfaces could be measured on the same sample.

All data presented in this paper have been obtained with patterned surfaces made of regular hexagonal arrays of cylindrical posts, as shown schematically in figure 2. The geometrical characteristics of the patterns are the heigth and the diameter of the posts (respectively *h* and *d*) and their spacing that we characterize by the center to center distance, *i* (see figure 2). We have chosen to fix *h*, and to vary in a systematic manner both *d* and *i*, in order to complement the data reported in Lamblet et al.[20] were the effect of *h* at fixed *i* and *d* was investigated.

We chose to fix the height of the pillars to *h*=2.2μm (except for one set of experiments were *h*=0.5μm), while the diameter and the spacing were respectively varied from 1.5μm to 8μm and from 3μm to 32μm. Examples of the patterned surfaces, are shown in figure 3.

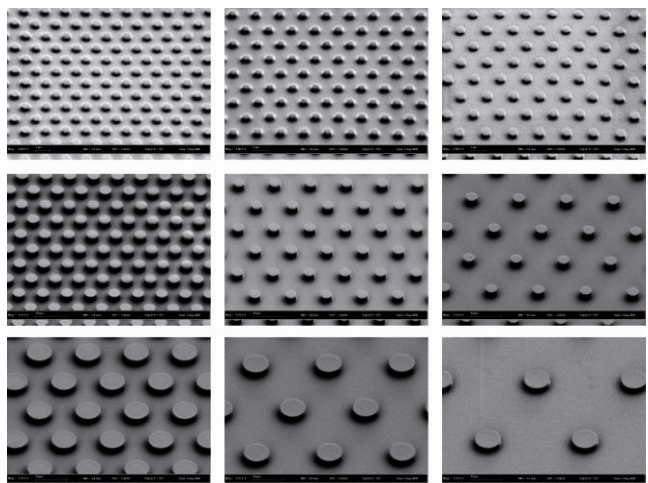

**Fig.3:** Examples of patterned surfaces. The height is 2.2μm for all pictures. For the first line, pillars diameter is 1.5μm and the spacing is from left to right 3, 3.5 and 4. For the second line, pillars diameter is 4μm and the spacing is from left to right 8, 12 and 16. For the third line, pillars diameter is 8μm and the spacing is from left to right 16, 24 and 32.

The acrylic adhesive was a commercial tape (3M600). Its width *b* is 19 mm, the thickness of the adhesive layer is 17 μm and that of the backing is 40μm. The storage modulus E'=0.02 MPa and the loss modulus and E''=0.005 MPa of the acrylic adhesive have been measured using dynamic shear experiments at 24°C and at a frequency of 0.1 Hz[22].

## Experimental results

A typical curve for the peel force as a function of the position of the peel front, either on a smooth or a textured part of the substrate is reported in figure 4.

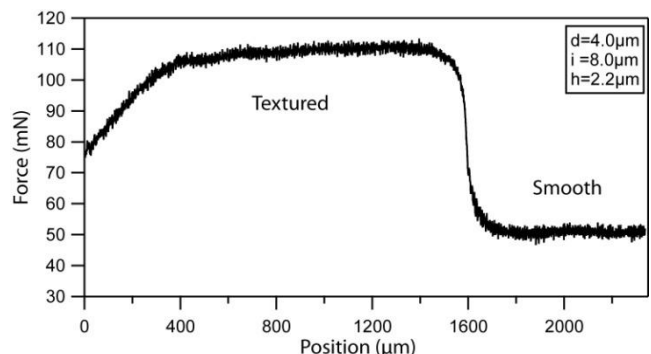

**Fig. 4**: Peel force versus position on textured and smooth interface.

The typical measured forces are in the range 30 to 150 mN. A neat increase of the adhesion force is obserevd when the peel front starts to move on the patterned surface, with a measured force typically twice that obtained on the smooth surface. Due to small differences in the curing or demolding process, the adhesion energy on the smooth part of the substrates was observed to slightly fluctuate and depend on the substrates. In order to isolate the effect of the patterning on the adhesion energy, we have systematically measured the peel force on the smooth part of each susbstrate, $F_s$, and used it to evaluate the adhesion enhancement due to patterning, from the difference between the peel force on the patterned part of the sample, $F_p$ and $F_s$. On the smooth surface, the peel energy per unit area, $G_s$, was deduced from the peel force $F_s$ by the well-known relation $G_s=F_s/b$, with *b* the width of the adhesive tape (cf. Fig.2).

On the patterned surface, the width of the patterned zone is smaller than that of the tape. The peel energy was then deduced taking into account the geometry of the contact (cf. Fig.2). On this part, both smooth and patterned zone contribute to the peel force. Assuming that each zone acts independently, the peel energy on patterned area is evaluated by:

$$G_p = \frac{F_p - F_s}{b_p} + \frac{F_s}{b} \quad (1)$$

We report below the relative increase of the peel energy $\Delta G/G_s=(G_p-G_s)/G_s$ as a function of the geometrical parameters of the texturation, for a fixed geometry: hexagonal arrays. Since a trivial effect of the patterning is to increase the surface of contact between the adhesive and the substrate (as we have chosen to work with the adhesive in full contact with the substrate), we present our results in terms of relative increase of the peel energy as a function of the relative increase of surface due to the texturation, $\Delta S/S_s$.

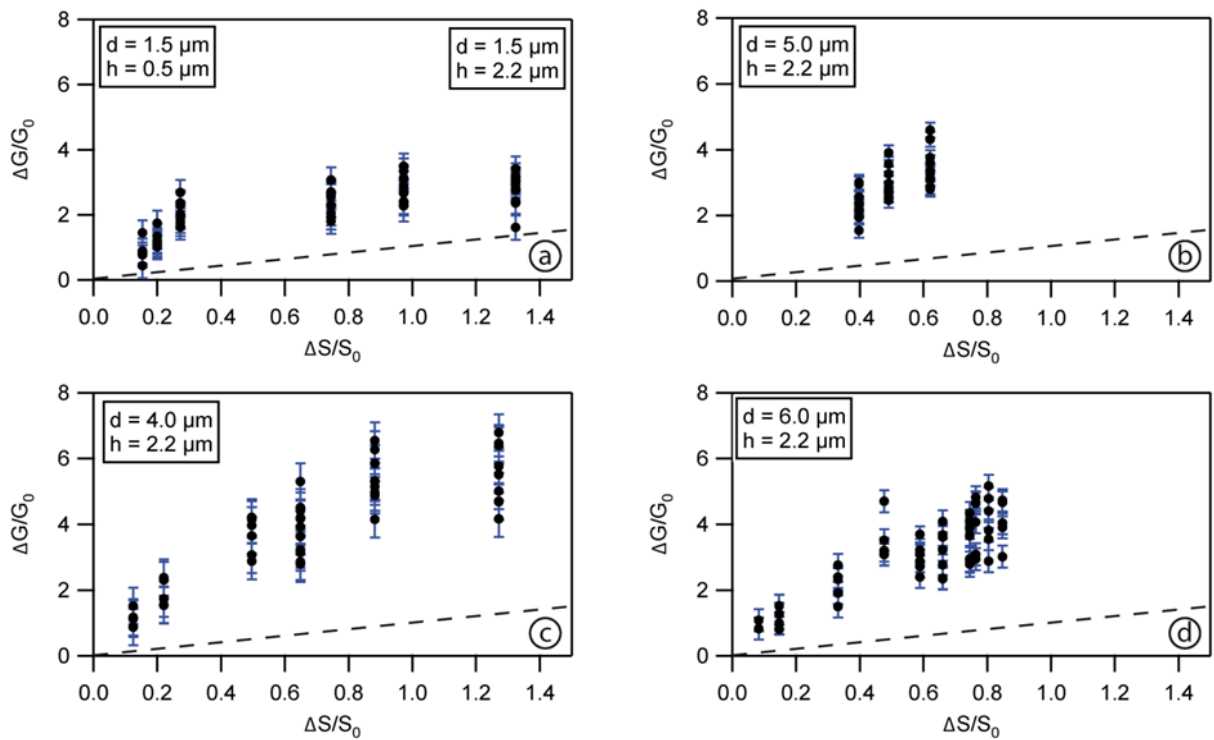

**Fig.5:** Evolution of the normalized enhancement of peel energy due to patterning as a function of the normalized increase of surface, for different diameters and spacing of the pillars. All data are obtained with pillars of height h=2.2μm, except the three series of data on the left hand part of curve 5a, for which the height has been decreased to 0.5 μm. The geometric characteristics of the pillars are all specified on each curve, and for each curve, the spacing *i* is varied to change the normalized increased area. The dashed line in each curve gives the contribution to the normalized peel energy resulting from the increase of area of contact.

For an hexagonal lattice, the number *N* of pillars per unit area is simply related to the size of the Wigner-Seitz cell in this geometry:

$$N = \frac{2}{\sqrt{3}i^2} \quad (2)$$

The surface increase per unit area is the lateral surface of one cylindrical pillar inside the Wigner-Seitz cell:

$$\frac{\Delta S}{S_S} = \frac{2\pi}{\sqrt{3}} \frac{dh}{i^2} = \frac{2\pi}{\sqrt{3}} \left(\frac{d}{i}\right)^2 \frac{h}{d} \quad (3)$$

The relative increase of surface $\Delta S/S_s$ is a complex combination of the aspect ratio of a pillar, $h/d$, and of the aspect ratio of the pattern, $d/i$. With the range of geometries explored, $\Delta S/S_s$ varies from 0.06 to 1.32. This parameter cannot be infinitely high since the spacing *i* is always higher than the diameter *d*. More precisely, with the used parameters for the lithography and the thickness *h* of the resist layer, actually, we were not able to have $i < d+1$ μm due to proximity effects inside the resin layer.

All results are summarized in Fig. 5. Each curve corresponds to given diameter and height except for curve a) for which data with two diameters and two heights are gathered. The spacing *i* is varied and allows one to span a certain range of normalized surface increase for each figure. Indeed, at fixed *d* and *h* increasing the spacing between pillars, decreases $\Delta S/S_s$ (see equation 3). Each symbol corresponds to one peel test. For one figure (fixed diameter and fixed height), the dispersion of the data has three main causes. First, different molds with the same nominal geometrical set of parameters were tested; and the lithography process may be responsible for a certain dispersion in the realizations for *d*, *h* and *i*. Second, successive molding were performed with a given mold. This again may lead to variability: one molding means a blend preparation, and some small variations in the composition of the reactive mixture can result in small variations of the Young modulus of the sample. Finally, the acrylic adhesive is hygroscopic, and the hygrometry was not fully controlled during these experiments.

All the curves in Fig. 5 clearly show that the adhesion energy increases with the relative surface increase due to the patterning. It is however worthwhile noticing that the reinforcement of the adhesion due to the texturation is not a simple trivial effect due to the increase of the contact area between the acrylic adhesive and the PDMS sample. This simple geometrical effect would lead to $\Delta G/G_s = \Delta S/S_s$, shown as the dash line in each sub-figure in figure 5. It is clear that the patterning is far more efficient than just increasing the surface area of contact between the substrate and the adhesive layer, and also that this adhesion enhancement cannot be simply described as a linear function of the surface increase.

Two regimes of adhesion enhancement can be identified. The first one, at small $\Delta S/S_s$ (large spacing), shows a linear variation of the enhancement of adhesion versus the surface increase. When the separation between the pillars becomes small, i.e. when $\Delta S/S_s$ becomes large, the relative increase of adhesion energy tends to level off and to become independent of the relative increase of surface. Comparing all data shown in all sub-figures 5, the absolute value of the plateau seems to depend on the pillars diameter, while the cross over between linear and saturated regimes of adhesion enhancement also depends on the pillars spacing. In the regime were the pillars are closed from each others, the level of adhesion enhancement at saturation increases with the diameter, with an earlier cross over for the smaller pillars.

A linear variation of the adhesion enhancement versus the normalized increased area of contact means that in this regime the pillars act independently of each other to enhance adhesion. It is then tempting to attribute the cross over to the saturated regime to a mechanical coupling of the pillars, when they become to close to each other. In reference 20, a first simple scaling description of the mechanical deformation of a single soft deformable pillar has been proposed, and shown to qualitatively account for the observed adhesion enhancement when the height of the pillars was varied at fixed diameter and spacing.

We present below a refined analysis of the mechanical response of the patterned soft substrate both for the situation were the pillars can be considered as independent of each other in their response to peel, and for the onset of coupling between pillars, leading to a saturated regime of adhesion enhancement whatever the relative increase in surface of contact with the adhesive.

## Model for the mechanical response of the patterned substrate

### Independent pillars: analytical model

As it was shown previously[20] that for low aspect ratio pillars, the adhesion anhancement associated to the patterning was essentially due to the elastic deformation of the substrate (pillars plus underlying PDMS film), we first propose to calculate this elastic energy of deformation as a function of the aspect ratio of the micropillars assuming that the pillars are acting independently of each other.

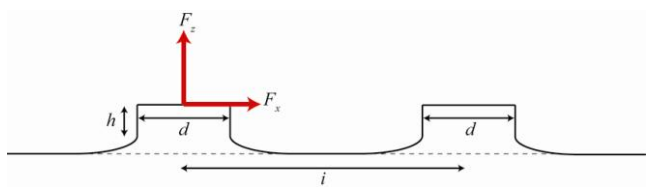

**Fig.6:** Definition of parameters used in the mechanical analysis.

Three contributions can be distinguished: each micropillar is bent and stretched, and the substrate itself can be deformed under the effect of the local peel force transmitted through the micropillar. From the video observations of the detachement of the patterned substrate from the adhesive, we observed that the final step of detachement is from the top of each pillar. The force at rupture on each pillar is called $F_p$. We do not know its exact orientation with respect to the substrate, due to the curvature of the adhesive and the deformation of the pillars. This force can anyhow always be split into a normal and a tangential components, $F_z$ and $F_x = F_z \tan(\alpha)$ respectively, with $\alpha$, the inclination angle of the peel force defined by $F_z = F_p \cos(\alpha)$. We have no obvious prediction for this angle but experimental observations tend to confort a approximative value of 45°. So, in a first approach, we shall consider that $F_x = F_z$. The geometrical parameters used in the modeling are summarized in figure 6.

The elastic energies $E_p$ stored by one cylindrical elastic micropillar both stretched and bend is classical and can, with the notations specified in figure 6, easily be written as[20]:

$$E_p = \frac{32}{3\pi} \frac{F_x^2}{E} \frac{h^3}{d^4} + \frac{2}{\pi} \frac{F_z^2}{E} \frac{h}{d^2} \qquad (4)$$

where $E$ is the Young modulus of the material. The first term in the right hand side in eq.(4) corresponds to the bending energy of the cylindrical elastic pillar submitted to the tangentiel force $F_x$ while the second one corresponds to the stretching energy of the pillar submitted to the normal force $F_z$.

The deformation energy of the underlying substrate is more delicate to evaluate. For a linear elastic and homogeneous substrate, the total mechanical energy can be expressed as[23]:

$$E_s = \frac{E}{(1-\nu^2)} \int_0^{d/2} \theta^2(\rho) d\rho \qquad (6)$$

where $\theta(\rho)$ is the Hankel transform of the normal surface displacement $u_z$ at radii values smaller than $d/2$, defined by:

$$\theta(\rho) = \frac{\partial}{\partial \rho} \int_0^\rho \frac{s \overline{u_z}(s)}{\sqrt{\rho^2 - s^2}} ds \qquad (7)$$

$\nu$ is the Poisson coefficient of the elastomer and is close to 0.5. The deformation field under a circular region submitted to the given normal force $F_z$, is [24]:

$$u_z(\rho) = \frac{6}{\pi^2} \frac{F_z}{Ed} \mathcal{E}\left(\frac{2\rho}{d}\right) \qquad (8)$$

where $\mathcal{E}(2\rho/d)$ is the complete elliptic integral of the second kind.

Combining eqs.((6),(7),(8)), the exact value of the stored mechanical energy can be calculated analytically (by using Maple V13.2):

$$E_s = \frac{4}{\pi^2} \frac{F_z^2}{dE} \qquad (9)$$

The total elastic energy per unit area of deformed patterned substrate under the effect of a force $F = F_x = F_z$ acting on each pillar can thus be simply calculated adding the contribution of each pillar, with the density of pillars (Eq. (2)):

$$E_{tot} = \frac{F^2}{E}\left(\frac{32}{3\pi}\frac{h^3}{d^4} + \frac{2}{\pi}\frac{h}{d^2} + \frac{4}{\pi^2}\frac{1}{d}\right)\frac{2}{\sqrt{3}i^2} \qquad (10)$$

Then, assuming that this elastic energy stored by the deformations of the pillars and of the underlying substrate is completely lost when the peel front passes, the peel energy becomes:

$$G_p = G_s\left(1 + \frac{\Delta S}{S_0}\right) + \frac{F^2}{E}\left(\frac{32}{3\pi}\frac{h^3}{d^4} + \frac{2}{\pi}\frac{h}{d^2} + \frac{4}{\pi^2}\frac{1}{d}\right)\frac{2}{\sqrt{3}i^2} \qquad (11)$$

The first terme in the right hand side term of equation (11) is the geometrical effect of the paterning, which increases the surface of contact between the substrate and the adhesive, corresponding to the dash line in figure 5. The second term in the right hand side term of equation (11) is the additional contribution due to the deformations of the elastic patterned substrate. This last term is proportional to the square of the applied force $F^2$. This force needs to be better identified in order to go further in the modeling.

It seems reasonable to identify the force $F_s$ as the critical pull-off force needed to separate the adhesive-pillar interface on the upper surface of each pillar, in view of the optical observations of the detachement between the adhesive and the patterned substrate. The question is then to precisely determine the pull-off force at rupture on the top of the pillar. The geometry of the top of the pillar is not perfectly a flat punch due to the deformability of pillars. It can be more or less approximated by a spherical cap. In the case of quasistatic separations, the adhesion force for an elastic contact between a spherical cap and a flat rigid medium has been calculated as:

$$F_c = \alpha \gamma \pi R \qquad (12)$$

with $R$ the radius of the sphere, and $\gamma$ the thermodynamic work of adhesion. The lower value of the numerical factor, $\alpha=3/2$, corresponds to the Johnson-Kendall-Roberts (JKR)[21] case, when adhesive forces act under the contact and soft deformable sphere. The highest value, $\alpha=2$, corresponds to the Derjaguin-Muller-Toporov (DMT)[25] situation, and is valid in the opposite limit of weak deformation of the sphere, and adhesion forces acting out of the contact. Intermediate cases have been worked out by Maugis[26]. The present situation with a small elastic sphere in contact with a soft adhesive, and probably non quasi-static sepation is more difficult to describe precisely. Since we do not know the detailled geometry of the top of the pillars, we chose as a simple approach to postulate a rupture criteria of the same kind than the JKR-DMT criterium:

$$F_c = \beta G_s \pi d \qquad (13)$$

where $\beta$ is a numerical factor and the chosen characteristic length is proportional to the diameter $d$ of pillars.

Replacing the value of the critical force in equation (11), by that in equation (13), one obtains a complete expression for the relative increase of peel energy:

$$\frac{\Delta G}{G_s} = 1 + \frac{\beta^2 G_s}{Ed}\frac{32}{3}\left(\frac{h}{d}\right)^2 + \frac{4d}{\pi h} + 2\frac{\Delta S}{S_s} \qquad (14)$$

All contributions related to the deformation of the patterned substrate, gathered in the second term inside the bracket of the right hand side of equation 14, appear inversly proportional to rthe diameter of the pillars. The bending contribution gives and additional square dependence in the aspect ration, h/d of the pillars, the substrate deformation only linearly with this aspect ration, while the stretching is of course independent of it. The relative increase of adhesion energy is thus predicted to increase linearly with the relative increase of surface for independent pillars, but with a slope bigger than the value one resulting from the simple geometrical effect of increased area of contact. If this analysis is correct, the comparison between equation (14) and experiments should allow one to determine the rupture criteria between the adhesive and the top of the pillars, all data being described with the unique fitting parameter $\beta$.

**Coupled pillars: numerical model**

When the distance between the pillars becomes small enough, the assumption of independent pillars can no longer hold. Experimentally, we have seen that the adhesion energy enhancement tends to saturate and to no longer depend on the relative increase of surface of contact (see the data in figure 5). The pillars become coupled. This coupling between pillars can a priori be mediated either by the adhesive itself, whose deformations should be affected by the proximity of the pillars, or by the elastic underlying PDMS layer. Since for low aspect ratio pillars, it was shown previously[11] that the enhancement of the adhesion was mainly due to the deformation of the PDMS (rigid patterned substrates were not efficient in promoting adhesion for low aspect ratio pillars), we have developed a numerical investigation of the coupling between the pillars due to the undelying substrate More precisely, the deformation field $u_z(r)$ given in equation (8) for the substrate contribution in the case of an isolated pillar needs be modified in order to take into account the coupling to the deformation fields associated to neighboring pillars. One needs determine the deformation field induced in the underlying substrate, taking into account the coupling between neighboring pillars.

Defining the distance between one pillar and the $n$-th neighboring pillar as $x_n=n.i$, the deformation field $u_z(\rho)$ in equation (8) can be rewritten as:

$$u_z(\rho) = \frac{6}{\pi^2}\frac{F}{Ed}\left[E\left(\frac{2\rho}{d}\right)\right.$$
$$\left.+ \sum_n \left[E\left(\frac{d}{2(\rho-x_n)}\right) - \left(1-\frac{d^2}{4(\rho-x_n)^2}\right)K\left(\frac{d}{2(\rho-x_n)}\right)\right]\right] \qquad (13)$$

The first term in the right hand side of the eq.(13) is the uncoupled part of $u_z(\rho)$ and the second term is the coupling part with $n$ neighbouring pillars[24]. This deformation field, eq.(6) and eq.(7) cannot be computed analytically to give the expression of the total elastic energy.

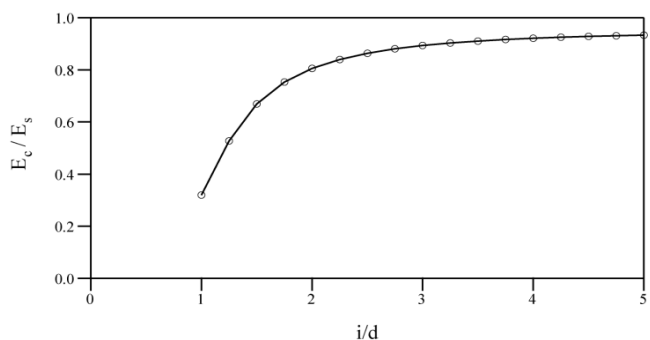

**Fig. 7:** Variation of the elastic energy of deformation including the coupling between neighboring pillars due to the deformation of the underlying substrate, $E_c$ normalized by the deformation energy for uncoupled pillars, $E_p$ (see eq.(9)) versus the horizontal aspect ratio of the pattern i/d.

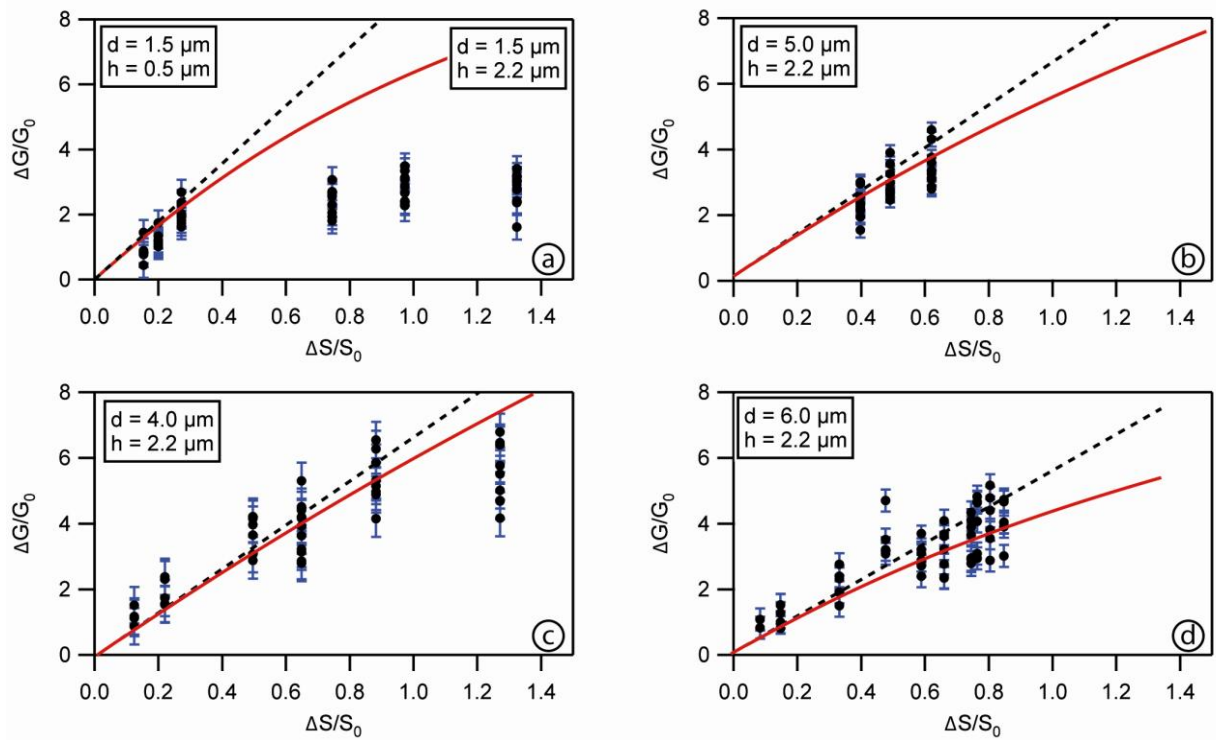

**Fig.8:** Comparison between calculated enhanced normalized peel energy and experiments (data yet shown in figure 5): analytical prediction for independent pillars (dash line) and numerical (full line) model for pillars coupled through the deformations of the elastic underlying substrate.

Due to the coupling through the substrate, the energy stored by the substrate under one pillar is not fully relaxed when this pillar detaches from the adhesive, and this leads to a saturation of the peel energy when the distance between pillars is decreased. The evolution of this energy $E_c$ normalized by the uncoupled energy $E_s$ (corresponding to eq.(9)) is reported as the function of horizontal aspect ratio $i/d$ of the pattern in figure 7.

This curve is independent of the exact value of the diameter $d$ of the pillars. It appears that the contribution of neighboring pillars remains negligible as long as the spacing $i$ is above three times the diameter $d$ and is approximatively divided by 3 when $i/d$ tends to 1.

## Comparison with experiments

The ensemble of curves in figure 8 provide a comparison between experiments and the prediction we have worked out both for uncoupled and for coupled pillars. In the model for independent pillars, the only adjustable parameter is the numerical coefficient β since all other parameters can be measured: $G_s$, is the peel energy on smooth interface, measured directly on each smooth zone of the substrates and equal to 4±1 J/m². We have not measured directly the Young modulus $E$ of the pillars, (which could be done, for example using colloidal probe AFM) and we have assumed that it was the same as in bulk, E=1.0±0.1 MPa (as measured by JKR technique).

Inserting these values in equation (14), it is possible to describe all data in the zone where pillars can be considered as independent (linear initial part of all curves in figure 8), whatever the diameter of the pillars, with as unique fitting parameter, the numerical factor β ~ 1±0,1. Moreover, as seen by the red curves in all sub-figures in figure 8, which represent the prediction of the numerical modelling, using the same set of parameters, the onset of coupling between pillars is well accounted for. This means that simple peel experiments on controlled micropatterned substrates can lead to the determination of a rupture criteria (determination of the numerical factor β) provided one uses a sufficient number of different geometries, in order to sense in details all regimes of mechanical response of the patterned substrate to peel solicitation.

## Conclusions

By investigating in a systematic manner the efficiency of surface micro-patterning in enhancing the adhesive strength at PDMS acrylic adhesive interfaces and comparing different diameter and spacing between pillars, we have shown that the peel energy increases first linearly with the surface increase resulting from the patterning. This linear increase is not indicative of a trivial effect of increaisng the area of contact between the adhesive and the patterned substrate, associated to the patterning, but can be quantitatively accounted for introducing the elastic deformation of the pillars, which is lost at detachment between the adhesive and the substrate, when the peel front passes. Two regimes of elastic deformation of the patterned substrate have been identified, depending on the aspect ratio of the pattern. When the cylindrical pillars used in the present study are far enough from each other (typically for distances between pillars larger than three times their diameter), the pillars behave independently of each other, and, for the relatively low aspect ratio pillars used, three kinds of elastic deformations contribute approximately equally to the peel energy: bending energy, stretching energy of each cylindrical pillar and deformation energy of the underlying elastic substrate. When the distance between pillars is decreased, the coupling between pillars, due to the coupling of the deformation field in the underlying elastic substrate leads to a saturation of the peel energy with the increase in surface of contact between the adhesive and the substrate.

The present set of experiments backs up the idea that the ability of the patterned surface to be deformed plays a crucial role in enhancing adhesion, similarly to what is seen, for example, in the case of animals like geckos.

Our investigations show that, by varying the size of the pattern, it is possible to tune the level of adhesion at PDMS acrylic adhesive interfaces. The enhancement of adhesion due to such patterning is purely elastic, and ruled first by the deformability of the patterned substrate, i.e. independent on the exact chemistry of the adhesive, and second by a rupture criteria on the top of the pillars which should remain sensitive to the chemistry of the adhesive.

We have shown that it was possible to extract this rupture criteria from the comparison between model and experiments, provided a sufficiently wide range of geometrical parameters of the patterned substrate can be used and compared.

Further experiments are presently underway to first directly access this rupture criteria, independently of the peel experiments, and second better control the geometry of the top of the pillars, so that the ways of controlling adhesion through micropatterning would be fully identified.

## Acknowledgements

C.P. greatly thanks E. Barthel for helpful discussions for the theoretical model. The authors thanks Y. Bardoux and T. Henry for experimental contributions and, D. Brunello and V. Klein for instrumental development. This work was supported by a "BQR Financier".

## Notes and references

*a Laboratoire de Physique des Solides, Univ Paris-Sud, UMR CNRS 8502, Bât. 510, Campus d'Orsay, F-91405 Orsay Cedex, FRANCE, E-mail: poulard@lps.u-psud.fr*